\documentclass[epsf,twocolumn,showpacs,preprintnumbers]{revtex4}
\usepackage{graphics,float}
\usepackage{graphicx}% Include figure files
\usepackage{dcolumn} % Align table columns on decimal point
\usepackage{bm}
\usepackage{epsfig}
\newcommand{\ucomma}[1]
{{\ooalign{#1\crcr\hidewidth\raise-.31ex\hbox{\scriptsize,}\hidewidth}}}
\begin{document}

\title{Mott Transition of MnO under Pressure: \\
   Comparison of Correlated Band Theories}
\author{Deepa Kasinathan,$^1$ J. Kune\v{s},$^{1,2}$ K. Koepernik,$^{3}$ 
Cristian V. Diaconu,$^4$ Richard L. Martin,$^4$ Ionu\ucomma t D.
  Prodan,$^5$ Gustavo E.
Scuseria,$^5$ Nicola Spaldin,$^6$ L. Petit,$^7$ T.C. Schulthess,$^7$
 and W. E. Pickett$^1$}
\affiliation{$^1$Department of Physics, University of California Davis,
  Davis, CA 95616}
\affiliation{$^2$Institute of Physics, ASCR, Cukrovarnick\'a 10,
  162 53 Praha 6, Czech Republic}
\affiliation{$^3$IFW Dresden, P.O. Box 270116, D-01171 Dresden, Germany}
%    Leibniz Institute of Solid State and Materials Research, Dresden, Germany
\affiliation{$^4$Theoretical Division, MSB269, Los Alamos National Laboratory
         Los Alamos NM 87545}
\affiliation{$^5$Department of Chemistry, Rice University, Houston TX 77005}
\affiliation{$^6$Materials Research Laboratory and Materials Department, 
         University of California Santa Barbara, Santa Barbara CA 93106}
\affiliation{$^7$Computer Science and Mathematics Division and Center for 
   Nanophase Materials Science, Oak Ridge National
   Laboratory, Oak Ridge TN 37831-6493}
\date{\today}
\begin{abstract}
The electronic structure, magnetic moment, and volume collapse 
of MnO under
pressure are obtained from four different correlated band theory methods; 
local density approximation + Hubbard U (LDA+U), pseudopotential 
self-interaction correction (pseudo-SIC),
the hybrid functional (combined local exchange plus Hartree-Fock 
exchange), and the local spin density SIC (SIC-LSD) method.
Each method treats correlation among the five Mn
$3d$ orbitals (per spin), including their hybridization
with three O $2p$ orbitals in the valence bands 
and their changes with pressure.  The focus is on comparison of
the methods for rocksalt MnO
(neglecting the observed transition to the NiAs structure in the 
90-100 GPa range).
Each method predicts a first-order
volume collapse, but with variation in the predicted volume and critical pressure.
Accompanying the volume collapse is a moment collapse, which for all methods is
from high-spin to low-spin ($\frac{5}{2}
\rightarrow \frac{1}{2}$), not to nonmagnetic as the simplest scenario would have.  
The specific manner in which the transition occurs varies considerably
among the methods: 
pseudo-SIC and SIC-LSD give insulator-to-metal, while LDA+U
gives insulator-to-insulator and the hybrid method gives an
insulator-to-semimetal transition.  
Projected densities of states above and
below the transition are presented for each of the methods and used to 
analyze the character of each transition.  In some cases the rhombohedral
symmetry of the antiferromagnetically ordered phase clearly influences the 
character of the transition.
\end{abstract}
                                                                                                                                                             
\maketitle
                                                                                                                                                             
\section{Introduction}

For fifty years the metal-insulator transition has been one of the
central themes\cite{MIT} of condensed matter physics.  The type we address here
does not involve spatial disorder nor change of the number of charge
carriers per cell; the competing tendencies arise solely from the kinetic
and potential energies in the Hamiltonian, favoring itineracy and 
localization respectively, and the many real-material complexities 
that arise.  The classic categorization is that of 
the Mott transition, treated in its most basic form with the
single-band Hubbard model.   Much has been
learned about this model, but there are very few
physical systems that are modeled faithfully by such a model.  Real
materials involve multiorbital atoms and thus extra internal degrees
of freedom, and an environment that is often very active and may
even react to the configuration of active sites.

MnO is a transition metal monoxide (TMO) with open $3d$ shell
that qualifies as one of the simpler
realizations of a prototypical, but real, Mott insulator.  It is, certainly,
a multiorbital system with the accompanying complexities, but the half-filled
$3d$ bands lead to a spherical, spin-only moment at ambient pressure.
Applying pressure to such a system leads to a number of possibilities,
including insulator-metal transition, moment reduction, volume collapse
if a first-order transition (electronic phase change)
occurs, and any of these may be
accompanied by a structural
phase transition, that is, a change in crystal symmetry. 
The $3d$ band width W of such a Mott
insulator is very susceptible to applied pressure, and is one of the
main determining factors of the strength of correlation effects. 

The ``closed subshell'' aspect makes MnO  
an atypical $3d$ monoxide,  
as shown for example by Saito {\it et al.}, who
compiled\cite{saito95a} effective parameters for this system from
spectroscopic information.  An effective intra-atomic Coulomb repulsion
energy as defined by them, for example, is roughly twice as large as
for the other $3d$ monoxides.
The complexity of this compound can be
considered in terms of the energy scales that are involved in the
electronic structure and magnetism of these oxides.  These include
the $3d$ bandwidth W, an intra-atomic Coulomb repulsion strength
U, an intra-atomic $d-d$ exchange energy (Hund's rule J, or exchange splitting
$\Delta_{ex}$), the crystal field splitting $\Delta_{cf} =
\varepsilon_{e_g} - \varepsilon_{t_{2g}}$, and the
charge transfer energy $\Delta_{ct} \equiv \varepsilon_d -
\varepsilon_p$ (the difference in mean Mn $3d$ and O $2p$ site energies).
In the magnetically ordered antiferromagnetic (AFM) state, there are
further symmetry lowering and ligand field subsplittings involving
$3d-2p$ hybridization.  All of
these scales change as the volume changes, making the pressure-driven Mott
transition a challenging phenomenon to describe.

\begin{figure}[H]
\begin{center}\includegraphics[%
  clip,
  width=0.4\textwidth,
  angle=-00]{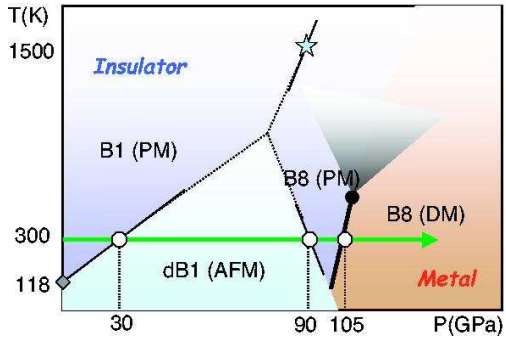}\end{center}
\caption{\label{MnOphasediagram} (Color online)
A schematic (conceptual) P-T phase diagram for MnO 
based on recent high pressure
work at Lawrence Livermore National Laboratory.\cite{PattersonMnO,YooMnO} 
The thick phase line signifies the first-order Mott
transition which simultaneously accompanies the loss of Mn magnetic
moment, a large volume collapse, and metalization.  This transition
should end at a critical point (solid circle).  The gray fan above the
critical point signifies a region of crossover to metallic
behavior at high temperature.  The star denotes earlier shock 
data,\cite{shock} and the diamond marks the ambient pressure N\'eel
temperature.  Only the distorted B1 (dB1) phase is
magnetically ordered.}
\end{figure}

Although the objective of the current paper is to compare methods within
the fcc (rocksalt) phase, it is useful first to recount what is known about the
Mott transition at this time.
The current experimental information, mostly at room temperature,
on the behavior of MnO under pressure
is summarized in Fig. \ref{MnOphasediagram}. 
Resistance measurements\cite{PattersonMnO} provided the first evidence
of the Mott transition in MnO near $100$ GPa.
Recent x-ray diffraction and emission spectroscopy measurements of
the crystal structure and magnetic moment by Yoo {\it et al.}\cite{YooMnO}
have clarified the behavior.  Around 90 GPa there is a structural
transformation from the distorted B1 (rocksalt) phase to the B8
(NiAs) structure.  This structure change is followed at 105 GPa by
the Mott transition, consisting of a simultaneous volume collapse and
moment collapse signifying a qualitative change in the electronic 
structure of the compound. 

On the theoretical side, little is known about how the Mott transition
occurs in a real multiband TMO in spite of the extensive studies of the
Mott transition in the single-band Hubbard model, which has a simple
spin-half moment at strong coupling and half-filling.  The numerous energy scales 
listed above, and the S=$\frac{5}{2}$ moment on Mn arising from the five
$3d$ electrons, allow many possibilities for how the moment might
disintegrate as the effective repulsion decreases.
The high pressure limit is clear: a nonmagnetic $3d-2p$ band metal in which kinetic
energy overwhelms potential energy.  This is the
competition studied in the (simplified) Hubbard model.  The
multiband nature has attracted little attention until recently, when
for example the question of  possible orbital-selective 
Mott transitions\cite{liebsch,koga} 
have aroused interest.  
One can imagine one scenario
of a cascade of moment reductions 
S = $\frac{5}{2} \rightarrow \frac{3}{2} \rightarrow
\frac{1}{2}$ before complete destruction of magnetism, as electrons use
their freedom to flip spins (as some competing energy overcomes Hund's
rule, for example).  In such a
scenario there is the question of which orbital flips its spin
at each spinflip, which involves a question of orbital selection and ordering.
At each flip the system loses exchange
(potential) energy while gaining kinetic energy (or correlation energy
through `singlet' formation). The manner in which kinetic
energy changes is difficult to estimate because subband involvement
means that there is no longer a single bandwidth W that is involved.
The increasing hybridization with O $2p$ states under pressure 
strongly affects the kinetic energy, directly and through 
superexchange (a kinetic energy effect). 

It is known that conventional band theory 
(local density approximation [LDA]) that does so well for so many 
materials gives poor results for $3d$ monoxides in many respects,
and some predictions are qualitatively incorrect ({\it viz.} no band
gap when there should be a large gap of several eV).  Thus even at the
density functional level (ground state energy, density, and magnetization)
some correlated approach is required.  In the past fifteen years several
approaches, which we refer to as correlated band theories, have been 
put forward, and each has had its successes in providing an improved
description of some aspects of correlated TMOs. 
Although commonly called mean-field approaches with which they share
many similarities, they are not mean-field treatments of any many-body
Hamiltonian.  Rather, they are energy functionals based on the
complete many-body Hamiltonian, which must then be approximated due to 
limited knowledge of the exchange-correlation functional.

In this paper, we provide a close comparison of certain results from
four such methods: 
full potential LDA+U, the hybrid exchange functional (HSE) approach, 
the self-interaction-corrected local spin-density method 
(SIC-LSD), and a nonlocal
pseudopotential-like variation of SIC (pseudo-SIC). 
Our main focus is to compare the predicted changes in energy, moment, and 
volume within the AFMII rocksalt phase of MnO.  To keep the comparison
manageable we confine our attention
to the rocksalt phase, since our emphasis is on comparison of methods
and not yet the ultimate but more daunting task of modeling 
structural changes that may precede, or accompany, the Mott transition.

\section{Structure and Symmetry}
Rocksalt MnO has an experimental
equilibrium lattice
constant $a_{\circ}$=4.45~\AA~(cubic cell volume V$_{\circ}$=88.1
~\AA$^3$).
Density functional theory, like Hartree-Fock theory, deals in its 
most straightforward form with ground state properties, {\it i.e.}
zero temperature.
The ground state is
known to be the AFMII phase in which 
$<111>$ layers have spins aligned,
and successive layers are antiparallel.  The resulting symmetry 
is rhombohedral, with Mn$\uparrow$ and Mn$\downarrow$ being 
distinct sites (although related through a translation + spin-flip
operation).  Thus, while most of the lore about transition metal monoxides is
based upon cubic symmetry of the Mn (and O) ion, in the ordered state
the electronic symmetry is reduced.  It is obvious that individual
wavefunctions will be impacted by this symmetry, {\it viz.} fourfold 
symmetry around the cubic axes is lost.  It has been emphasized by 
Massidda {\it et al.}\cite{massidda} that zone-integrated, and even
spin-integrated, quantities show the effects of this symmetry lowering;
for example, Born effective charges lose their cubic symmetry.
Since this issue arises in the 
interpretation of our results, we provide some background here.

In cubic symmetry the Mn $3d$ states split into the irreducible 
representations (irreps) denoted by $t_{2g}$ and $e_g$.  Rhombohedral
site symmetry results in the three irreducible representations $a_g$,
$e_{g,1}$, and $e_{g,2}$, the latter two being two fold degenerate. 
The coordinate rotation from cubic to rhombohedral (superscript $c$
and $r$ respectively) is, with a specific choice for the orientation
of the $x$ and $y$ axes in the rhombohedral system,
\[
\left( \begin{array}{c}
           x^{r} \\
           y^{r} \\
           z^{r} \\ \end{array} \right) = \left( \begin{array}{ccc}
           \frac{1}{\sqrt{6}} & \frac{1}{\sqrt{6}} & -\frac{\sqrt{2}}{\sqrt{3}} \\
           -\frac{1}{\sqrt{2}} & \frac{1}{\sqrt{2}} & 0 \\
           \frac{1}{\sqrt{3}}  & \frac{1}{\sqrt{3}} & \frac{1}{\sqrt{3}}\\
           \end{array} \right) \left( \begin{array}{c}
                                      x^{c} \\
                      y^{c} \\
                                      z^{c} \\  \end{array} \right)
\]
Applying this rotation of coordinates gives the $3d$ orbitals in the
rhombohedral frame in terms of those in the cubic frame ($d_{z^2} \equiv
d_{3z^2-r^2}$):
\begin{eqnarray}
d_{xy}^r &=&\frac{1}{\sqrt{3}} \left( d_{xz}^c - d_{yz}^c - d_{x^2-y^2}^c \right)\\
d_{yz}^r &=&\frac{1}{\sqrt{6}}(d_{yz}^c - d_{xz}^c)- \sqrt{\frac{2}{3}}d_{x^2-y^2}^c\\
d_{xz}^r &=&\frac{\sqrt{2}}{3}d_{xy}^c - \frac{1}{3\sqrt{2}}(d_{xz}^c + d_{yz}^c)
           - \frac{\sqrt{2}}{\sqrt{3}}d_{z^2}^c \\
d_{x^2-y^2}^r &=&-\frac{1}{3}(d_{xz}^c + d_{yz}^c + \frac{2}{3}d_{xy}^c)
           - \frac{1}{\sqrt{3}}d_{z^2}^c\\
d_{z^2}^r &=&\frac{1}{\sqrt{3}}(d_{xy}^c + d_{yz}^c + d_{xz}^c).
\end{eqnarray}

In rhombohedral coordinates it is useful to categorize the $3d$ orbitals in
terms of their orbital angular momentum projections along the rhombohedral
axis: $d_{z^2}^r$ $\leftrightarrow$ $m_{\ell}=0$; 
$d_{xz}^r, d_{yz}^r \leftrightarrow m_{\ell}=\pm 1$;
$d_{xy}^r, d_{x^2-y^2}^r \leftrightarrow m_{\ell}=\pm 2$.
It is easy to see that $|m_{\ell}|$ specifies groups of states that only transform into
combinations of themselves under trigonal  point group operations.

Note that the unique $a_g$ symmetry state in rhombohedral
coordinates is the fully symmetric combination of the cubic $t_{2g}$ states. 
The other two irreps are both $e_g$ doublets.  While $|m_{\ell}| = 1$ and 
$|m_{\ell}| = 2$ form representations of these irreps, if there are components
of the crystal field that are not diagonal in the L=(2,$m_{\ell}$) basis,
these states will mix.  Then each of the resulting (orthonormal) irreps
$e_{g,1}$, and $e_{g,2}$ will contain both $|m_{\ell}| = 1$ and
$|m_{\ell}| = 2$ components.  Such mixing does occur in MnO and complicates
the symmetry characterization of the $3d$ states.

\section{Methods}

\subsection{LDA Calculations}
For LDA band structure plot (Fig. \ref{LDAbands}) we used version 5.20 of the 
full-potential local orbital band structure
method (FPLO\cite{Koe99}). Relativistic effects were 
incorporated on a
scalar-relativistic level.
We used a single numerical basis set for the core states (Mn $1s2s2p$
and O $1s$) and a double
numerical basis set for the valence sector including two $4s$ and $3d$
radial functions, and
one $4p$ radial function, for Mn, and two $2s$ and $2p$ radial functions,
and one $3d$
radial function, for O. The semi core states (Mn $3s3p$) are treated as
valence states with a single numerical radial function per
$nl$-shell.  The local density exchange-correlation functional PW92 of
Perdew and Wang\cite{PW92} was used.

\subsection{LDA+U Method}
The LDA+U approach of including correlation effects is to (1)  
identify the correlated orbital, $3d$ in this case, (2) 
augment the LDA energy functional with a
Hubbard-like term (Coulomb repulsion U) and Hund's (exchange J) energy 
between like  spins, (3) subtract off a spin-dependent average of this 
interaction energy to keep from double-counting repulsions (once in 
LDA fashion, once in this U term), and (4) include the correlated 
orbital occupation numbers in the self-consistency procedure, which leads
to an orbital-dependent Hartree-Fock-like potential acting on the correlated 
orbitals.  The addition to the energy functional has the schematic form
\begin{eqnarray}
E_U=\frac{1}{2}\sum_{m\sigma \neq m'\sigma'} (U - J\delta_{\sigma\sigma'})
     [n_{m\sigma} n_{m'\sigma'} -\bar{n}_{\sigma} \bar{n}_{\sigma'}].
\label{ldau}
\end{eqnarray}

We actually use the coordinate-system independent form of
LSDA+U \cite{Esc03,Czy94,Ani97} implemented in FPLO,\cite{Koe99} which 
leads to four $m$ indices on $U$ and $J$ which for simplicity
have not been displayed
(nor has the full off-diagonal form of the occupation matrices 
$n_{m m'\sigma}$).  This treatment of the on-site interactions $U$ and $J$ 
incorporates on-site correlation effects in the Mn 3d-shell. We have used the
so called `atomic-limit' (strong local moment) form of the 
double-counting correction, the last term in Eq. \ref{ldau}.  This form is
appropriate for the high-spin state, but it is less obviously so for the
low-spin state that is found at reduced volumes.  The
Slater parameters were chosen according to $U=F_0=5.5~\mathrm{eV}$,
$J=\frac{1}{14}(F_2+F_4)=1~\mathrm{eV}$ and $F_2/F_4=8/5$. \\

The shape of the basis orbitals has been optimized yielding
a sufficient accuracy of the total  energy over the range of geometries
considered in this work. The $\mathbf{k}$-integrals are performed via
the tetrahedron method with an irreducible mesh corresponding to 1728
($12^3$) points in the full Brillouin zone.

\subsection{SIC-LSD Method}
The SIC-LSD method addresses the unphysical self-interaction in the
LDA treatment of localized states.  Itinerant states, being spread over
space without finite density in any given region, do not experience this
self-interaction within the LDA treatment. Should there be localized
states, confined to some region and giving a finite density, they will suffer an
unphysical self-interaction in the LSD method.  This issue then clearly arises in
the itinerant-localized transition in MnO and other correlated systems.
The basic premise of the SIC-LSD method is that localized
electrons should experience a different potential from that of itinerant 
electrons,\cite{Cowan67,Lindgren71,PZ81} analogous to that of an atomic state
whose self-interaction must be removed.  Then electrons on the surrounding atoms
are allowed to accommodate self-consistently.  This distinction of localized versus
itinerant state is addressed in SIC-LSD by extending the energy functional
in the form
\begin{eqnarray}
E^{SIC-LSD}&=&E^{LSD} - \sum_{\alpha}^{occ.}\delta_{\alpha}^{SIC}, \\ \nonumber
 \delta_{\alpha}^{SIC}&=&U[n_{\alpha}] + E_{xc}^{LSD}[n_{\alpha}].
\end{eqnarray}
Here $U[n]$ represent the Hartree (classical Coulomb) energy of a
density $n(r)$. 
The self-Coulomb energy $U[n_{\alpha}]$ and self-exchange-correlation
$E_{xc}^{LSD}[n_{\alpha}]$ energies are subtracted off for each localized state
$\psi_{\alpha}$ with density $n_{\alpha}$.  Whether states are localized
or not (with non-zero, respectively zero self-interaction) is determined
by minimization of this functional, allowing localized as well as
itinerant states $\psi_{\alpha}$.  Since the correction vanishes for 
itinerant states, the sum finally includes only the self-consistently
localized states.  The localized and itinerant states are expanded in the
same basis set, and minimization becomes a process of optimizing the
coefficients in the expansion of the states (as other band structure
methods do, except that Bloch character is imposed in other 
methods).  
The implementation of Temmerman and collaborators\cite{walter} 
used here incorporates the atomic-sphere approximation (ASA) of the linear
muffin-tin orbital electronic structure method\cite{LMTO} (LMTO) in the
tight-binding representation.\cite{lmtotb} 
Further details can be found in Ref. \onlinecite{walter2}.

\subsection{ pseudo-SIC Method}
The large computing requirements (compared to LDA) of the SIC-LSD method,
even for materials with small unit cells, has led to an alternative approach
\cite{Vogel96}, in which the self-interaction part of the Kohn-Sham
potential is approximated by a non-local, atomic-like contribution included within
the pseudopotential construction. The original implementation of this
scheme has given important improvements over
LSDA results for non-magnetic II-VI and III-V semiconductors, but was not applicable to
metals or to magnetic and highly-correlated systems where there is a coexistence of
strongly localized and hybridized electron charges.

The pseudopotential self-interaction corrected calculations presented here were performed
using the recently developed ``pseudo-SIC'' method of Filippetti and Spaldin
\cite{Filippetti03}. This pseudo-SIC approach represents a
compromise between the fully self-consistent implementations of Svane {\it et al.}
\cite{walter,svane1990} and the alternative method of
Vogel et al. \cite{Vogel96}, in that the SIC calculated for the atom
(as in Ref. \cite{PZ81}) is scaled by the electron occupation numbers calculated
self-consistently within the crystal environment. This allows 
the SIC coming from localized,
hybridized, or completely itinerant electrons to be discriminated, 
and permits the treatment
of metallic as well as insulating compounds, 
with minimal computational overhead beyond the LSDA.
In this pseudo-SIC procedure, the orbital SIC potential is taken from the
isolated neutral atom and included in the crystal potential in
terms of a nonlocal projector, similar in form to the nonlocal part of the
pseudopotential. The Bloch wave functions are projected onto the basis of
the pseudo-atomic orbitals, then, for each projection, the potential 
acting on the Bloch state is
corrected by an amount corresponding to the atomic SIC potential.
Note that, within this formalism, a physically meaningful energy functional
which is related to the Kohn-Sham equations by a variational principle is
not available. However, a suitable expression for the total energy
functional
was formulated in Ref. \cite{Filippetti03} and shown to yield
structures
in good agreement with experiment. We use this functional here to calculate
the bondlength dependence of the total energy. We have used ultra-soft 
pseudopotentials with an energy cutoff of 35 Ry.
An $8^3$ Monkhorst-Pack grid was used for $k$-point sampling. The
low-spin and high-spin solutions were obtained by setting initial magnetization to
5$\mu_B$ or 1$\mu_B$ respectively. \\

\subsection{Hybrid Functional Method}
The hybrid-exchange DFT approximation mixes a fraction of the exact, non-local,
exchange interaction (which uses the Hartree-Fock [HF] expression)
with the local, or semi-local,
exchange energy of the LDA or the generalized gradient approximation
(GGA).  The PBE0 functional takes the form:
\begin{eqnarray}
 E_{xc}  = a E_x^{HF} + (1-a) E_x ^{PBE} + E_c ^{PBE}
\end{eqnarray}
where $E_x^{HF}$ and $E_x^{PBE}$ are the exchange and the PBE
GGA functionals.  The mixing parameter $a$ = 1/4 was determined 
via perturbation theory
[21] and $E_c^{PBE}$ is the PBE correlation energy.

  In this work we use the hybrid method recently developed by Heyd, Scuseria and
Enzerhof (HSE)).\cite{Heyd}  It is based upon the PBE0 functional, but employs a
screened, short-range (SR) Hartree-Fock (HF)
exchange instead of the full exact exchange, which results in 
a more efficient evaluation
for small band gap systems. In this approach, the Coulomb operator is
split into short-range (SR) and long-range (LR) components respectively
\begin{equation}
\frac{1}{r} = \frac{1-erf(\omega r)}{r} + \frac
              {erf(\omega r)}{r},
\end{equation}
where $\omega$ is a
parameter that can be adjusted for numerical or formal convenience. 

The expression for the HSE exchange-correlation energy is
\begin{eqnarray}
E_{xc}&=&aE_{x}^{HF,SR}(\omega) + (1-a)E_{x}^{PBE,SR}(\omega) \\ \nonumber
      & & + E_{x}^{PBE,LR}(\omega) + E_{c}^{PBE},
\end{eqnarray}
where E$_{x}^{HF,SR}$($\omega$) is the SR HF exchange computed for the SR
part of the Coulomb potential,  E$_{x}^{PBE,SR}$($\omega$) and 
E$_{x}^{PBE,LR}$($\omega$) are the SR and the LR components of the PBE 
exchange, respectively.  The cited papers should be 
consulted for further details.
The HSE functional has been found to yield results in good agreement with
experiment for a wide range of solids and molecules \cite{gust1,gust2}.

This functional is implemented in the development version 
of the Gaussian quantum chemistry 
package \cite{Frisch}.  We use the Towler basis\cite{Towler1,Karen,Towler2}
of Gaussian functions for our basis set.
It consists of a [20s12p5d/5s4p2d] contraction for Mn and
a [14s6p/4s3p] basis for O, optimized for HF 
studies on MnO. A pruned grid for numerical integration with 99 radial
shells and 590 angular points per shell was used. The ${k}$-space was sampled
with a $16^3$ mesh. 
Low spin and high spin 
antiferromagnetic initial guesses were obtained using the 
crystal field approach 
by patching the density matrix obtained from diagonalization of the Harris
functional \cite{harris} with the density matrices obtained from calculation
on ions in the appropriate ligand field \cite{kudin,prodan}.

\section{Previous Electronic Structure Studies}
The origin, and the proper description, of the moments and the band gaps
in transition metal monoxides have a long history.
The earliest question centered on the connection between the antiferromagnetic
(AFM) order and the insulating behavior.  Slater's band picture\cite{slater}
could account in a one-electron manner for a gap arising from AFM order,
whereas Mott's picture of correlation-induced insulating behavior\cite{mott}
was a many-body viewpoint with insulating behavior not connected to the
magnetic order.  The proper general picture in these monoxides arose from
studies of transport above the N\'eel temperature and with introduction of
defects, giving them the designation as Mott insulators.  

Much progress on the understanding of MnO and the other monoxides came from
early studies using LDA.  While understanding that LDA does not address the
strong correlation aspect of the electronic structure, Mattheiss\cite{mattheiss}
and Terakura {\it et al.}\cite{terakura} quantified the degree and effects
of $3d-2p$ interactions, and pointed out the strong effect of magnetic 
ordering on the band structure.  More recently, Pask and collaborators\cite{pask}
have studied the structural properties, and the rhombohedral distortion,
with LDA and GGA approximations.  The symmetry lowering and resulting structure
is described well, and in addition they found that AFM ordering results in 
significant charge anisotropy.  Effects of AFM order were further probed
by Posternak {\it et al.} by calculating and analyzing maximally localized
Wannier functions for the occupied states.\cite{posternak}

The application of correlation corrections in MnO already has a colorful
history.  The first work, by Svane and Gunnarsson\cite{svane1990} and by
Szotek {\it et al.},\cite{szotek} was in the application of the SIC-LSD
method.  The former pair correctly obtained that MnO, FeO, CoO, NiO, and CuO
are AFM insulators, while VO is a metal.  They calculated a gap of 4 eV for
MnO.  Szotek {\it et al.} used a fairly different implementation of
the SIC-LSD approach but find a similar gap (3.6 eV).  Their $3d$ states
lay about 6 eV below the center of the $2p$ bands, although hybridization
was still clearly present. 
In this same time frame, Anisimov, Zaanen, and
Andersen introduced\cite{Ani91} the LDA+U method with application to the
transition metal monoxides.  They obtained a band gap of 3.5 eV but few
other results on MnO were reported.   

Kotani implemented\cite{kotani1,kotani2,kotani3} the ``DFT exact-exchange'' method of 
Talman and Shadwick\cite{talman} to crystal calculations.  This method 
consists of taking the Fock expression for the exchange energy in the DFT
functional, then performing a Kohn-Sham solution (minimization), giving a
local exchange potential (``optimized effective potential'').  
In Kotani's results for MnO, the Mn $e_g$ and
$t_{2g}$ bands form very narrow (almost atomic-like) bands between the 
occupied O $2p$ bands and the conduction bands.  Takahashi and 
Igarashi\cite{igarashi} proposed starting from the Hartree-Fock exchange
and adding correlation from a local, three-body scattering viewpoint.
Their corrections were built on a parametrized tight-binding representation,
and they obtained small self-energy corrections for MnO, much smaller
than they obtained for the other transition metal monoxides.

The effective potential approach used by Kotani was extended by
Solovyev and Terakura\cite{solovyev} in an unconventional way.  They
obtained an effective potential using the criterion that it had to
reproduce the spin-wave spectrum, {\it i.e.} that it had to describe the
magnetic interactions correctly.  They found clear differences when 
comparing to the LDA+U and the optimized effective potential results,
and discussed limitations of the one-electron band method itself.

More recently, Savrasov and Kotliar applied a dynamical extension\cite{dmft} of the
LDA+U method (dynamical mean field theory) to MnO and NiO.  Being a self-energy
method, this is not really a correlated band theory.  For the properties they
calculated (band gap, effective charges, dielectric constant, optic phonon
frequencies) the dynamical results are similar to the LDA+U results and
differ considerably from LDA values.

  Even though hybrid-exchange DFT applications to solids are still in their infancy,
there have been two previous studies of MnO.  The first, by 
Bredow and Gerson,\cite{bredow}
utilized the B3LYP hybrid functional.  Unlike the LDA and GGA, they found B3LYP
provided an excellent band gap for MnO.  More recently, Franchini {\it et al.}
have examined MnO in more detail using the PBE0 approximation.\cite{franchini}  They also
found a gap,
lattice constant and density of states in quite good agreement with
experiment.  In particular, the distorted dB1 rhombohedral structure was
determined to be the minimum energy geometry, in agreement with experiment.
Neither the B3LYP nor the PBE0 approximation can be applied to the metallic
side of the transition of interest here.  For that, we must turn to the
screened hybrid-exchange of HSE.

Therefore, while there has been thorough LDA studies of MnO and a variety
of approaches to treatment of the correlation problem, nearly all of these 
have considered only ambient pressure or small variations of volume near
zero pressure.  The work described in the following
sections focuses on testing the four different correlated band methods from
ambient conditions to high pressures, through the volume collapse regime, to
see whether some basic foundation can be laid for the understanding and
theoretical description of pressure-driven Mott transitions in real materials.

\section{Results}
Our principal results revolve around the first-order transition, high$\rightarrow$low
volume which is also high$\rightarrow$low moment in nature.
For convenience we use from here on the specific volume $v\equiv V/V_0$, the volume
referenced to the experimental zero-pressure volume.
The equations of state have been fitted for both high volume and low volume
phases for each computational method, and the resulting constants are
presented and analyzed below.  

\subsection{Baseline: LDA Bands and Equation of State}
\begin{table}[tb]
\begin{center}
\begin{tabular}{|c|c|c|c|c|}
\hline
       &      &          &         &          \tabularnewline
       & FPLO & Gaussian & PW-USSP & LMTO-ASA \tabularnewline
       &      &          &         &          \tabularnewline
\hline
$a/a_o$& 0.97 &   0.97   &  0.96   &   0.97   \tabularnewline
B (GPa)&  170 &    196   &   169   &    205   \tabularnewline
Gap (eV)& 0.72 &  1.13   &  0.92   &   1.04   \tabularnewline
Moment ($\mu_B)$& 4.52 & 4.53 & 4.42 & 4.42 \tabularnewline
\hline
\end{tabular}
\end{center}
\caption{\label{lda_table} Local density approximation results
obtained from the four codes used in this work, for the equilibrium
lattice constant relative to the experimental value ($a/a_o$),
the bulk modulus B,
the energy gap, and Mn moment.  The experimental values
for B fall in the range 142-160 GPa (see caption to Table
\ref{data_table}).  All calculations are
carried out for the antiferromagnetic II phase of MnO in the cubic
(NaCl) structure.  Differences are due to (1) differing exchange-correlation
functions that are used, (2) approximations made in algorithms to solve
the Kohn-Sham equations and evaluate the energy, (3) basis set quality,
and (4) for the moment, the operational definition of the ``Mn moment''
differs.  The notation follows. FPLO: full potential local orbital method
with Perdew-Wang (1992) exchange-correlation functional. Gaussian: Gaussian
local orbital code, Slater exchange ($\alpha$=2/3) and Vosko-Wilk-Nusair
correlation. PW-USSP: planewave basis using ultrasoft pseudopotentials,
with the Perdew-Zunger exchange-correlation functional.
LMTO-ASA: linear muffin-tin orbital code in the atomic sphere
approximation, using the Perdew-Zunger exchange-correlation functional.}
\end{table}

\begin{figure}[tb]
\begin{center}\includegraphics[%
  clip,
  width=0.4\textwidth,
  angle=-90]{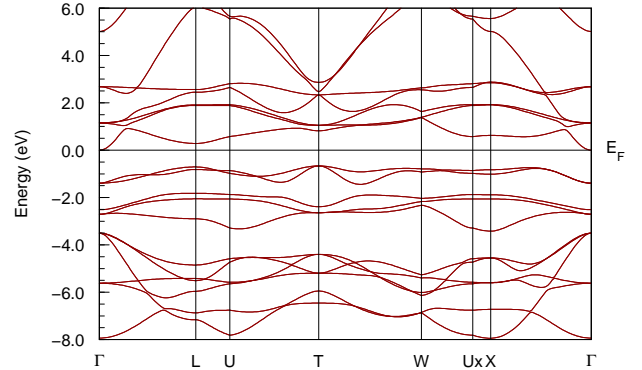}\end{center}
\caption{\label{LDAbands} (Color online)
LDA band structure of AFM MnO along rhombohedral symmetry
lines, calculated with the FPLO 
method\cite{Koe99}, with horizontal line (``Fermi level'')
placed at the top of the gap. 
The $\Gamma$-T lies along the rhombohedral axis, while $\Gamma$-L lies
in the basal plane.  The O $2p$ bands lie in the -8 eV to -3.5 eV range,
with the majority Mn $3d$ bands just above (-3 eV to -1 eV).  The 
five minority $3d$ bands are just above the gap.  Note the small
mass, free-electron-like band that lies below the unoccupied $3d$ 
bands at the $\Gamma$ point.
}
\end{figure}

In Table \ref{lda_table} we illustrate the magnitude of variation of
three properties calculated from the four codes used here, as an indication
of what size of differences should be given meaning in properties 
presented below.  Before the method-specific beyond-LDA corrections are
applied, there are differences in the bandgap (0.95$\pm$0.2 eV) due to the 
algorithms applied in the codes, to the basis quality, and because different
LDA functionals are used.  It is evident that these variations have very
little effect on the calculated equilibrium volume and minor effect on 
the calculated Mn moment (4.47$\pm$0.05 $\mu_B$).  Such differences will
not affect the comparisons of the results given here, at the level of
precision of interest at this time.

The LDA band structure of AFM MnO is shown in Fig.
\ref{LDAbands} as the reference point for the following calculations.
There is a band gap of $\sim 0.7$ eV.
The five bands immediately below the gap are the majority Mn $3d$ bands,
those lying below are the O $2p$ bands.  
The charge transfer energy mentioned in the Introduction is $\Delta_{ct}
= \varepsilon_d -\varepsilon_p$ = 6 eV, and the exchange splitting is
$\approx$3.5 eV.
It is tempting to interpret the
3+2 separation of occupied $3d$ states as $t_{2g} + e_g$, but the
rhombohedral symmetry renders such a characterization approximate.
The five bands above the gap
are primarily the minority Mn $3d$ bands.  However, a
free-electron-like band at $\Gamma$ lies lower in energy than the $3d$ bands,
but disperses upward rapidly, so over most of the zone the lowest
conduction band is Mn $3d$ and the gap is 1 eV.
The presence of the non-$3d$ band does
complicate the interpretation of the band gap for some of the correlated
methods, presented below.

The behavior of MnO under compression within GGA has been
given earlier by Cohen, Mazin, and Isaak.\cite{cohen}
They obtained an equilibrium volume 2\% higher, and bulk modulus 13\%
smaller, than measured.   Pressure studies including extensive
structural relaxation have also been provided by Fang {\it et al.}\cite{fang}
Their structural relaxations make their study more relevant (within the
restrictions of GGA) but also make comparison with our (structurally
restricted) results impossible.

\subsection{Energetics and Equation of State}
\begin{table}[H]
\begin{center}
\begin{tabular}{|c|c|c|c|c|c|}
\hline
&
&
&
\tabularnewline
& GGA&
LDA+U&
HSE&
pseudo&
SIC \tabularnewline
& &
& exchange
& -SIC 
& -LSD
\tabularnewline
\hline
\hline
$v_0$ & 1.02&
$0.93$&
$0.99$&
$1.09$&
$1.04$ \tabularnewline
\hline
$v_h$ & 0.70 &
0.66  &
0.60  &
0.86  &
0.64 \tabularnewline
\hline
$v_l $ & 0.62 &
0.61  &
0.55  &
0.73  &
0.52  \tabularnewline
\hline
$\Delta v$  & 0.08 &
 0.05 &
 0.05 &
 0.13 &
 0.12\tabularnewline
\hline
\hline
$B_h$ & 196 &
192  &
187  &
138  &
159 \tabularnewline
\hline
$B_l$ & - &
195  &
224  &
230  &
67 \tabularnewline
\hline
\hline
$B'_h$ &  3.9 &
3.2 &
3.3 &
3.6 &
3.3\tabularnewline
\hline
$B'_l$ & - &
3.6 &
4.0 &
3.5 &
4.7\tabularnewline
\hline
\hline
$P_{c}$ & 149 &
123 &
241 &
56  &
204\tabularnewline
\hline
\end{tabular}
\end{center}
\caption{\label{data_table}
Quantities obtained from fits to
the Murnaghan equation
of state for the various functionals, except for the GGA column, which are
taken from Ref. \cite{cohen}. $v_0$ is the calculated equilibrium volume,
B and B' are the
bulk modulus (in GPa) and its pressure derivative. $v_{h}, v_{l}$ are 
the calculated volumes of the high and low pressure phases, respectively,
at the critical pressure $P_c$ (in GPa). 
$\Delta v$ is the amount of volume 
collapse that occurs at
the transition pressure $P_{c}$.  All volumes are referred to the
experimental equilibrium volume.  The experimental values are $B$=142-160
GPa, $B'\approx 4$; see Zhang\cite{zhang} and references therein.}
\end{table}

\begin{figure}[tb]
\begin{center}\includegraphics[%
  clip,
  width=0.5\textwidth,height=0.5\textwidth,
  angle=-90]{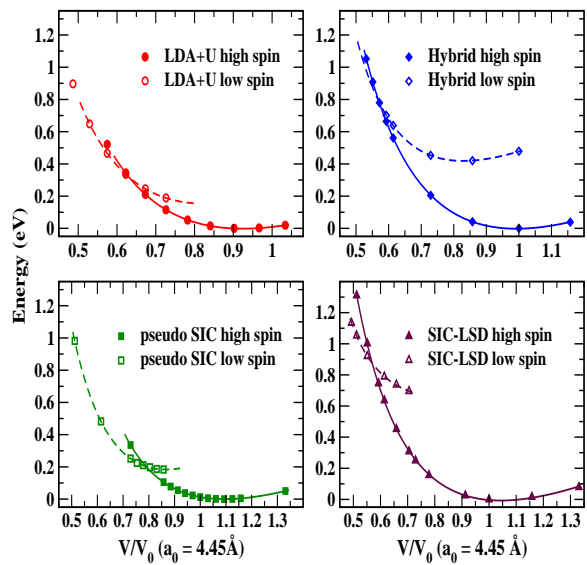}\end{center}
\caption{\label{EvsV}(Color online)
The calculated total energy/MnO versus volumes for the
various functionals for the AFMII rocksalt phase, referred to zero at
their equilibrium volume. The filled symbols
denote the calculated energies of the large volume, high spin configuration and 
the open symbols
denote the calculated energies of the small volume, low spin configuration. 
The continuous and
dashed lines are the least square fitted curves to the Murnaghan equation 
of state for high and low spin configurations respectively.}
\end{figure}

The equation of state (EOS) energy vs. volume curves 
for the various functionals are collected
in Fig. \ref{EvsV}. For each correlated band method a large volume, high-spin state
and a small volume, low-spin state are obtained.  The analysis to
obtain the first-order volume collapse transition was done as follows.
For each high volume and low volume phase separately, an  
EOS function $E_{h,l}(V)$ was determined ($h,l$=high, low)
by a fit to the Murnaghan equation.
Both fits give minima, with the most relevant one being for the high spin
phase and being the predicted equilibrium volume $V_0^{th}$.
The pressure is obtained from the volume derivative of the EOS,
which is inverted to get $V(P)$.  Equating the enthalpies
$E[V(P)] + PV(P)$ of the two phases gives the critical pressure $P_c$.
The volumes at this pressure then give the volume collapse 
$\Delta V$ = $V_h(P_c)
-V_l(P_c)$.  

The various quantities for all four computational schemes
are given in Table \ref{data_table}, along with the uncorrelated 
results of Cohen {\it et al.}\cite{cohen}
Not surprisingly given the other differences that will be discussed,
there are substantial variations among the critical pressures and 
related quantities. 
Particularly noticeable already in the EOS  is the result 
that the energy difference between the low-spin energy minimum and the
high-spin one is $\sim$0.2 eV for the LDA+U and pseudo-SIC methods,
while the HSE method gives roughly twice the energy difference (0.4 eV),
and the SIC-LSD method gives roughly 0.6 eV. 

We now mention other noteworthy features of the calculated data in Table
\ref{data_table}.\\  
(1) The predicted equilibrium volume from the LDA+U method
is the smallest of the four methods ($v_0^{th}$=0.93), thus overbinding.  
The HSE value is almost indistinguishable from the observed value,
while the SIC-LSD and pseudo-SIC methods give
underbinding ($v_0^{th}$ = 1.04, 1.09 respectively).\\
(2) The pseudo-SIC method predicts the transition to occur at a relatively
small volume reduction ($v^h$=0.86); the other methods give the onset of
transition at $v$=0.63$\pm$0.03. \\
(3) The critical pressure P$_{c}$=56 GPa predicted by pseudo-SIC is smallest
of the methods.  P$_c$ in LDA+U (123GPa) is comparable to that of LDA; those
of SIC-LSD and HSE are higher (204, 241 GPa, respectively).\\
(4) The SIC-LSD method predicts a transition to a low volume phase that is 
much softer than the high volume phase, a phenomenon that is extremely
unusual in practice but not disallowed.  There are two possible sources of
this difference: (i) in
SIC-LSD the system becomes completely LDA-like in the low volume phase
whereas in the other methods the $3d$ states are still correlated, or
(ii) the LMTO-ASA method involves approximations that pose limitations in
accuracy.\cite{lmtolimits}\\
(5) The values of $B$ in the large volume phase vary although not anomalously
so, given the differences discussed just above. \\  
(6) The values of $B'$ in the high volume phase are reasonably similar across
the methods: $B' = 3.4\pm0.2$.  The variation in the collapsed phase is
greater.

\subsection{Magnetic Moment}
\begin{figure}[tb]
\begin{center}\includegraphics[%
  clip,
  width=0.5\textwidth,height=0.5\textwidth,
  angle=-90]{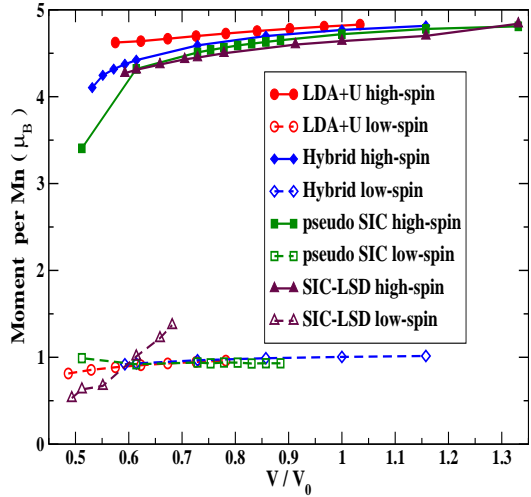}\end{center}
\caption{\label{MvsV} (Color online)
Calculated values of the moment (shown as symbols) 
on each Mn site as a function
of volume for the various functionals. All the methods predict a distinct
collapse (first order) of magnetic moment with decrease in volume. At 
large volumes, the high spin state with S = 5/2 (single occupancy of each of the 
$3d$ orbitals) is realized while the low spin state with S = 1/2 is
favored for smaller volumes. Note: the computational methods calculate 
the `Mn moment' in inequivalent ways, so differences of less than
$0.1 \mu_B$ may have no
significance.}
\end{figure}

The moment collapse behavior of each method is collected in 
Fig. \ref{MvsV}.  
For comparison, the GGA result presented by Cohen
{\it et al.}\cite{cohen} was a moment collapse from 3.4$\mu_B$ to 
1.3$\mu_B$ at the volume given in Table \ref{data_table}.
On the broadest level, the predictions for the Mn moment show {\it remarkable
similarity}.  At low pressure, all methods of course give the high-spin
S=$\frac{5}{2}$ configuration of the Mn$^{2+}$ ion, with the local
moment being reduced slightly
from 5$\mu_B$ by $3d-2p$ mixing.  This electronic phase persists over a
substantial volume reduction,  giving way 
in all cases to an S=$\frac{1}{2}$ state,
{\it not} the nonmagnetic $S=0$ result that might naively be anticipated.
Three methods give a stable moment very near 1$\mu_B$ over a 
range of volumes.  The SIC-LSD method (which in the collapsed phase is
simply LDA) is alone in giving a varying moment, one that reduces 
from 1.4 $\mu_B$ at $v=0.68$ down to 0.5 $\mu_B$ at $v=0.48.$

One difference between the methods lies in how soon the
low-spin state becomes metastable, {\it i.e.} when it is possible 
to obtain that state self-consistently, as opposed to when it becomes 
the stable solution (which was discussed in the EOS subsection).  
The state is obtained already
at ambient volume in the HSE method; the pseudo-SIC method obtains the
low-spin state just below $v=0.80$; for the LDA+U method, it 
was not followed above $v=0.68$.  It should be emphasized however that
no concerted measures were taken to try to follow all solutions
to the limit of their stability.

\begin{figure}[tb]
\begin{center}\includegraphics[%
  clip,
  width=0.5\textwidth,height=0.5\textwidth,
  angle=-90]{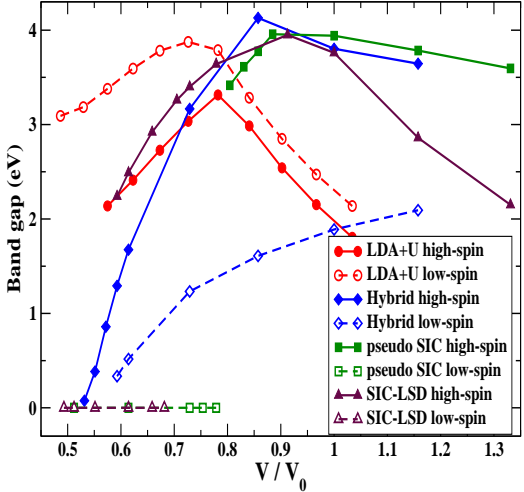}\end{center}
\caption{\label{gapvsa}(Color online)
Calculated band gap as a function of volume; the lines are simply a
guide to the eye. 
For LDA+U, the band gap increases with decrease in volume for the
high spin state, but decreases with volume in the low spin state. 
For the HSE calculations, the band gap monotonously decreases with volume
whilst pseudo-SIC shows a first order jump in going from high to low spin 
state. At very low volumes, where the low spin configuration is preferred
LDA+U gives a substantial gap and is still an insulator, while HSE and
pseudo-SIC calculations converge to a metallic solution.}
\end{figure}

\subsection{Fundamental Band Gap}
In Fig. \ref{gapvsa} the calculated band gap of both high-spin and low-spin
states are shown for all methods.  Here the behavior differs considerably
between the methods, in part because at certain volumes 
the gap lies between different bands
for some of the methods.  At ambient pressure the pseudo-SIC and HSE methods obtain a gap 
of 3.5-4 eV, while that for SIC-LSD is 2.9 eV, and that of the LDA+U method 
is even lower, less than 2 eV.  Experimental values lie in the 3.8-4.2 eV
range.  Referring to Fig. 2, it can be observed
that the large volume gap depends on the position of the majority $3d$ states
with respect to the free-electron-like band, {\it i.e.} it is not the
$3d-3d$ Mott gap.  Both
of the former approaches show only a slight increase as pressure is applied,
reaching a maximum around $v=0.76$ where a band crossing results in a
decreasing gap from that point.  For
pseudo-SIC, there is an almost immediate collapse to a metallic low spin state.
Within HSE, MnO collapses to the low-spin state 
at a volume $v=0.55$, coincident with the closing of the gap in the
low-spin state.  The SIC-LSD and LDA+U gaps, smaller initially,
show a much stronger increase with pressure, and also incur the band crossover
that leads to decrease of the gap (within the high-spin state).

%For the pseudo-SIC, SIC-LSD, and HSE methods, 
%the collapse is to a metallic state
%(zero gap).  
%The LDA+U transition has a distinctive character: the low-spin state has a 
%larger gap than the high-spin state, and there does not seem to be a metallic
%state nearby. 

\section{Analysis of the Transition}
In this section we analyze the character of the states just above and just
below the Mott transition, as predicted by each of the methods.  Due to
the differing capabilities of the codes, the quantities used for analysis 
will not be identical in all cases.  In the Figs. 6-9 we present for uniformity
the DOS in the high volume phase at the equilibrium volume ($a_0$) and in
the collapsed phase at $a = 0.85 a_0$ ($v$=0.6).  Note (from Table I) that 
this specific volume does not correspond to any specific feature in the
phase diagram for any method, although it lies in the general neighborhood
of the volume at the collapse.  Changes within the collapsed phase are
continuous, however, so the plots at $a = 0.85 a_0$ are representative
of the collapsed phase. 

\subsection{LDA+U}
\begin{figure}[tb]
\begin{center}\includegraphics[%
  clip,
  width=7.8cm,
  angle=-00]{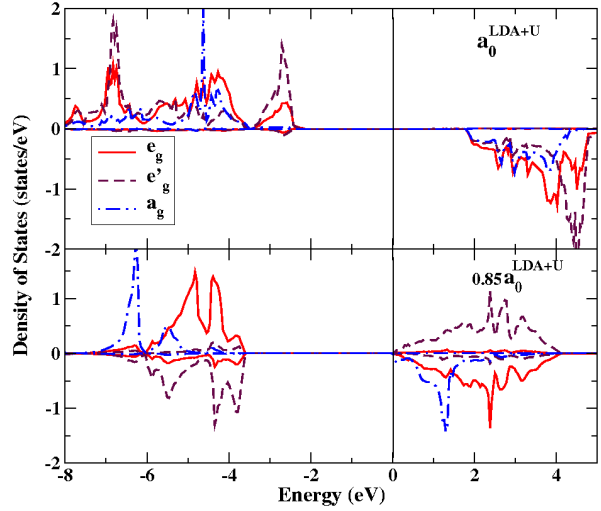}\end{center}
\caption{\label{ldaudos}(Color online)
Projected DOS onto symmetrized Mn 3$d$ orbitals
in the rhombohedral AFMII rocksalt
phase using the LDA+U method.  Top panel: High spin solution at the
LDA+U equilibrium volume.  Bottom panel: Low spin solution at 60\% of the
LDA+U equilibrium volume.
The $a_{g}$ orbital is the $3z^2-r^2$
oriented along the rhombohedral axis, other symmetries are described in
the text.  The overriding feature is the spin-reversal of the $m=\pm 1$
$e'_g$ orbitals between the two volumes.}
\end{figure}

The projected DOSs (PDOSs) in Fig. \ref{ldaudos} refer to projections 
onto Mn $3d$ orbitals, with the $z$-axis being the rhombohedral axis, the $a_g$
$3z^2-r^2$ ($|m|$=0) state; the $e'_g$ pair \{$xz,yz$\} ($|m|$=1); and the 
$e_g$ pair \{$x^2-y^2,xy$\} ($|m|$=2).  Because the two $e_g$ representations
have the same symmetry, they can mix and the actual combinations $e_{g,1}, e_{g,2}$
are orthogonal linear combinations of $e_g, e'_g$ which depend on
interactions.  For the LDA+U results, however, there is little mixing of 
the $e_g, e'_g$ pairs.
The character of the transition is simple to
describe: the $e'_g$ pair ($|m|$=1 with respect to the rhombohedral axis)
simply flips its spin.

This S=$\frac{1}{2}$ state is unexpected and quite unusual.  First, each
$3d$ orbital is still singly occupied, verified by plotting the charge
density on the Mn ion and finding it just as spherical as for the high-spin
state.  Second, each $3d$ orbital is essentially fully spin-polarized, 
with the configuration
being $a_g\uparrow e_g\uparrow e'_g\downarrow$.  A plot of the spin density\cite{deepa}
reveals the unanticipated strong anisotropy with nodal character, 
characteristic of spin-up $m=0$ and $|m|$=2
orbitals, and spin-down $|m|$=1 orbitals (in the rhombohedral frame).
Third, it makes this transition with essentially zero
change in the gap, which is 3.5 eV.  The band structure changes completely,
however, so the close similarity of the gaps on either side is accidental.

\begin{figure}[tb]
\begin{center}\includegraphics[%
  clip,
  width=7.8cm,
  angle=-00]{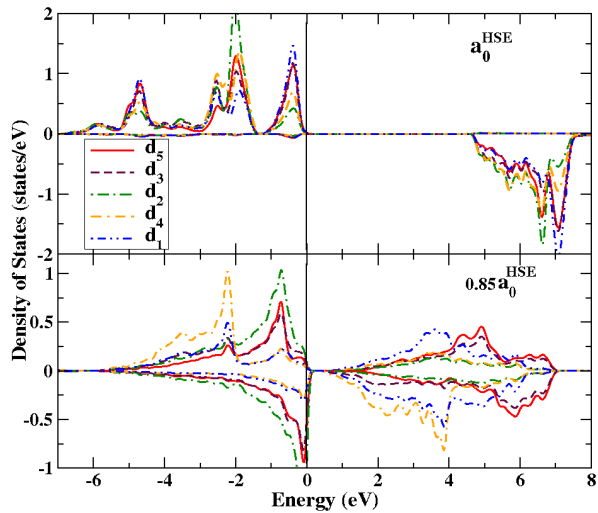}\end{center}
\caption{\label{hsedos}(Color online)
Spin- and orbital-projected DOS from the HSE (hybrid-exchange)
method.  Orbitals are expressed in a coordinate system that is specific to
the code, neither cubic nor rhombohedral.  In the 
collapsed phase two orbitals are (nearly) doubly occupied, 
and the net moment arises
from the single occupation of the orbital labeled $d_4$.  In a cubic
representation the 
configuration can be specified as 
$t^5_{2g} = t^3_{2g\uparrow} + t^2_{2g\downarrow}$.
At the volume $v$=0.55 of onset of the collapsed phase (see Table I),
the gap is essentially zero. }
\end{figure}

\subsection{HSE Method}
In the high
volume phase the distribution and overall width of the occupied $3d$ states,
shown in Fig. \ref{hsedos},
is similar to that of the LDA+U method (previous subsection).  The gap
is larger, as discussed earlier.
The collapsed phase shows new characteristics.  
The gap collapses from 4 eV at $a_0$ to essentially zero at
the onset of the collapsed phase at $v$=0.55, making this an
insulator-to-semimetal transition.  The metallic phase then evolves
continuously as the pressure is increased beyond $P_c$.
The $3d$ configuration can be characterized as $t^5_{2g} = 
t^3_{2g\uparrow} + t^2_{2g\downarrow}$, resulting in a
moment of 1$\mu_B$.  The corresponding spin density is strongly 
anisotropic, although in a different manner than is the case for LDA+U.

\subsection{pseudo-SIC Method}
\begin{figure}[tb]
\begin{center}\includegraphics[%
  clip,
  width=7.8cm,
  angle=-00]{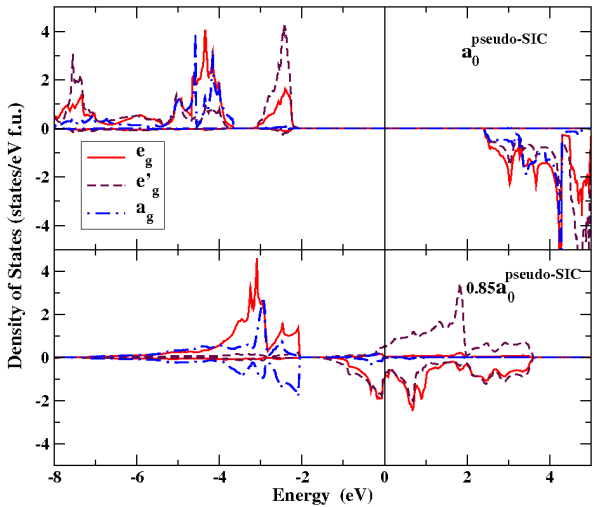}\end{center}
\caption{\label{pSICdos} (Color online)
Spin- and symmetry-projected DOS of the high
spin (upper panel) and low spin (lower panel) states resulting from
the pseudo-SIC method.  In the low-spin state, the $a_g$ orbital is
unpolarized due to occupation by both spin directions,
and shows little exchange splitting.  One of the $e_g$-symmetry
pairs, here called $e_g'$, is unpolarized due to being unoccupied in
both spin directions. It is the one labeled $e_g$ that is spin split
across the Fermi level and is responsible for the moment.
}
\end{figure}

The spin-decomposed spectrum from this method is shown in Fig. 
\ref{pSICdos}, symmetry-projected as done above for the LDA+U method.
The PDOSs in the high-spin state are quite similar to those given
by the LDA+U method.  The transition could hardly be more different,
however.  The gap collapses in an insulator-to-good-metal character,
the Fermi level lying within both majority and minority bands.
The majority bands are the $e'_g$ pair ($|m|$=1) and are only slightly
occupied.  In the minority bands both $e_g$ and $e'_g$ are roughly
quarter filled.  The reason this solution is (at least locally) stable
seems clear: E$_F$ falls in a deep valley in the minority DOS.

In the collapsed, low-spin state, the $a_g$ orbital of both spins is occupied, and 
the majority $e_g$ pair is also fully occupied.  This results in a
configuration that can be characterized roughly as
$a_g^1\uparrow e_g^2\uparrow; a_g^1\downarrow e_g^{0.5}\downarrow 
(e'_g)^{0.5}\downarrow$, giving spin 
$\frac{3}{2}-\frac{1}{2}-\frac{1}{2}=\frac{1}{2}.$
Thus the fact that the same moment is found in the low-spin state as was found
with the LDA+U and pseudo-SIC
methods seems accidental, because in those methods the energy gap required
integer moment whereas the pseudo-SIC solution is firmly metallic.  It is in fact close
to half metallic, which accounts for the near-integer moment.
In pseudo-SIC, the $a_g$ orbital is unpolarized (spin-paired),
the $e_g$ pair \{$xz,yz$\} ($|m|$=1) is positively polarized, and
the $e'_g$ pair \{$x^2-y^2,xy$\} ($|m|$=2) is negatively polarized
but to a smaller degree.

\begin{figure}[tb]
\begin{center}\includegraphics[%
  clip,
  width=7.8cm,
  angle=-00]{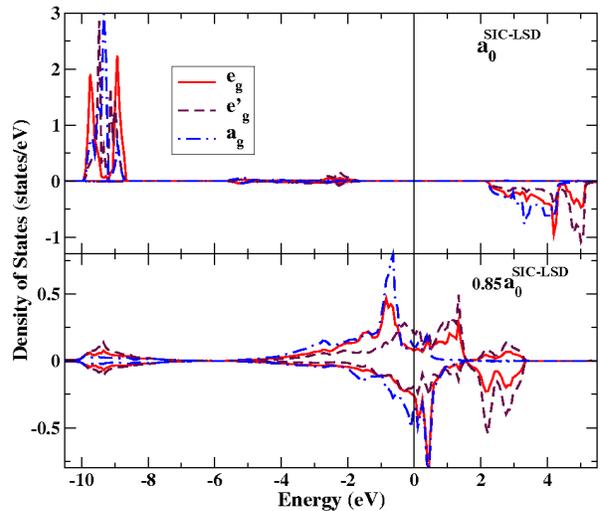}\end{center}
\caption{\label{LSDSICdos}(Color online)
Spin- and symmetry-projected DOS of the high
spin (upper panel) and low spin (lower panel) states given by
the SIC-LSD method.  Orbital characters are expressed in the rhombohedral
coordinate system.  In the high-spin state, the majority $3d$ states are
are centered 6 eV below the center of the occupied $2p$ bands, resulting
in little hybridization and very narrow bands.  
The exchange splitting of the $3d$ states is 
about 13 eV, providing an ``effective U'' from this method.  The
collapsed moment phase is representative of a band (LSD) ferromagnet.
}
\end{figure}

\subsection{SIC-LSD Method}
This method give much more tightly bound $3d$ state in the high-spin
state than the other methods.  An associated feature is that the 
majority-minority splitting, the ``effective
U''$\equiv U^{SIC}$ plus the exchange splitting 13 eV, more than twice as
la large as used in the LDA+U method (both in this paper and elsewhere).  
Note that in the SIC-LSD method $U^{SIC}$ 
is a true Slater self-Coulomb integral, whereas in the LDA+U method the value 
of $U$ represents the (somewhat screened) Coulomb interaction between a $3d$
electron and an {\it additional} $3d$ electron, so agreement between the
two is not expected.  Nevertheless the difference is striking.
All five majority $3d$ states are localized, leading
to the self-interaction potential that binds them.  The majority $3d$ states
lie 6 eV below the center of the $2p$ bands and hybridize very weakly, 
which accounts for the very narrow, almost core-like $3d$ bands.

In the collapsed phase, there are no localized states and the usual LSDA
results re-emerges.  All $3d$ states make some contribution to the moment,
but the strongest contribution arises from the $a_g$ orbital.  One might
argue in the LDA+U result that the net moment arises from the $a_g$
orbital (with the moments arising from the $e_g$ and $e'_g$ orbital
canceling).  However, the electronic structure in the collapsed phase of
this SIC-LSD method is very different from that of the LDA+U method.

\section{Discussion and Summary}
The results of four correlated band methods -- LDA+U, SIC-LSD, pseudo-SIC,
and HSE -- have been compared for the equation of state (for both normal
and collapsed phases), electronic structure
(including $3d$ configuration), and the Mn moment under pressure.  In order to
make the comparison as straightforward as possible, the crystal structure
was kept cubic (rocksalt).  To compare seriously with experiment, one must
account for the coupling of the AFM order to the structure, because this
results in a substantial rhombohedral distortion of the lattice.\cite{fang}
Then structural transitions (particularly to the B8 phase) must 
also be considered.

The large volume, high-spin phases are qualitatively the same for the 
various functionals: AFM with a fully-polarized $3d^5$ configuration.  
Due to the large charge transfer energy, the configuration remains $d^5$
at all volumes studied here.  The
predicted equations of state (which give the equilibrium volume, bulk 
modulus, and its pressure derivative) show rather strong variation,
suggesting that the extension of the Mn  $3d$ and O $2p$ functions, or
their hybridization, differ substantially even in this large volume
phase.  Of course, since the functionals are different, any given density
would lead to different energies.

Under pressure, the gap initially increases (all methods give this behavior), and
the system suffers a first-order transition (isostructural, by constraint) 
to a collapsed phase where hybridization must be correspondingly stronger.  Uniformly
among the methods, the moment collapse reflects an S=$\frac{5}{2}$ to
S=$\frac{1}{2}$ transition, rigorously so for the LDA+U method
for which the collapsed phase retains a gap, 
and less rigorously 
so for the the other methods where the collapsed phase is
metallic.                                                                                 
It is remarkable that none of the methods gives a collapse to a
nonmagnetic state, which probably would be the most common expectation.  

This S=$\frac{5}{2}\rightarrow \frac{1}{2}$ moment collapse is related 
in some cases to 
the local symmetry of the Mn $3d$ orbitals in the AFM phase
(being most obvious for the
LDA+U results of Fig. \ref{ldaudos}).
The symmetry is $a_g + e_g + e'_g$, {\it i.e.} a singlet and two doublets
per spin direction.  Without further symmetry-breaking (orbital ordering)
an S=$\frac{5}{2}\rightarrow \frac{3}{2}$ transition requires a single
spin flip, which could only be the $a_g$ spin.  However, the $a_g$ state
is more tightly bound than as least one of the two doublets both for
LDA+U (Fig. \ref{ldaudos}) and for pseudo-SIC (Fig. \ref{pSICdos}).
In LDA+U the $e_g$ doublet flips its spin, while in pseudo-SIC the $a_g$
singlet flips its spin leaving the minority $e_g$ and $e'_g$ 
doublets partially occupied and therefore metallic.
This symmetry-related behavior
depends of course on the magnetic ordering that gives rise to the
(electronic) rhombohedral symmetry.  Above the N\'eel
temperature, the moment collapse at the Mott transition may proceed
differently because the Mn moment would lie at a site of `cubic' symmetry
(a dynamic treatment could include the effect of short range spin correlations).

Our study 
provides some of the first detailed information on how magnetic 
moments in a real material may
begin to disintegrate without vanishing identically, at or near 
a Mott transition, when correlation is taken 
into account.  It is accepted that dynamic processes will be required for
a truly realistic picture of the Mott (insulator-to-metal) transition.
However, a moment collapse between two insulating phases (as described
here by two of the methods) may be described reasonably by a correlated
band (static) approach.

Beyond this similar amount of moment collapse, the four functionals give
substantially different collapsed phases: differences in the Mn $3d$ 
magnetic configuration (although all remain $d^5$)
and differences in conducting versus insulating
behavior.  It is not surprising therefore that the collapsed-phase equations of
state differ considerably between the methods.

The differences in predictions can be traced, in principle, to the 
different ways in which exchange and correlation are corrected with
respect to LDA.  One clear shortcoming of LDA is in the local approximation
to the exchange energy.  The HSE method deals with this
problem directly, by using 25\% Hartree-Fock
exchange.  The self-interaction of the SIC-LSD method is largely
a self-exchange energy correction, subtracting out the spurious 
self-Coulomb energy that occurs in the Hartree functional if an
orbital chooses to localize.  A self-(local density) correlation correction
is also included in SIC-LSD.  The pseudo-SIC method includes the same
correction if applied to an atom, but in a crystal the pseudo-SIC 
energy correction and change in potential takes a substantially different
form, as the difference in predictions reflects.  The LDA+U method is
rather different in this respect: it specifically does not subtract out
any self-interaction (although it is sometimes discussed in this way).
In the form Eq. \ref{ldau} of LDA+U correction, the second term is 
simply an LDA-like average of the first term.  The on-site Coulomb
repulsion is treated
Hartree-Fock like, leading to an orbital-dependent, occupation-dependent
potential.  Each method has its own strengths, and each is only
an anticipated improvement on LDA toward a better, more general functional.
It is expected that more details of the results may be published
separately by the respective practitioners.

\section{Acknowledgments}
D.K., J.K. and W.E.P. acknowledge support from Department of Energy
grant DE-FG03-01ER45876.  R.L.M. and C.V.D. thank the DOE BES heavy element 
chemistry program and LANL LDRD for support.  The work at Rice University was 
supported by DOE Grant DE-FG02-01ER15232 and the Welch Foundation.
We acknowledges important interactions within, and some financial support from, 
the Department of Energy's
Stewardship Science Academic Alliances Program.
L.P. and T.C.S. were supported by the Office of Basic Sciences, U.S.
Department of Energy.
N.S. was supported by the National Science Foundation's Division of
Materials Research Information Technology Research program,
grant No. DMR-0312407, and made use of MRL Central Facilities
supported by the National Science Foundation Award No. DMR-05-20415.
This collaboration was stimulated by, and supported by, DOE's
Computational Materials Science Network.

\end{document}